 \documentclass[3p,times,12pt]{elsarticle}
 \makeatletter
\def\ps@pprintTitle{%
 \let\@oddhead\@empty
 \let\@evenhead\@empty
 \let\@oddfoot\@empty
 \let\@evenfoot\@empty
}
\makeatother

\usepackage{etoolbox}
\patchcmd{\emailauthor}{(#2)}{}{}{}
\patchcmd{\urlauthor}{(#2)}{}{}{}

\usepackage{hyperref}
\hypersetup{
    colorlinks=true,       % false: boxed links; true: colored links
    linkcolor=blue,        % color of internal links
    citecolor=blue,        % color of links to bibliography
    filecolor=magenta,     % color of file links
    urlcolor=cyan          % color of external links
}

\usepackage{amssymb}
\usepackage{amsmath}
\begin{document}

\begin{frontmatter}



 \title{Efficient Simulation of Deformable Particles}
 %%\tnotetext[label1]{}
 \author{Roshan Maharana}

 %\fntext[label2]{}
 %\cortext[cor1]{}
 \affiliation{organization={Tata Institute of Fundamental Research},%Department and Organization
           % addressline={Gopanpally Tanda}, 
            city={Hyderabad},
            postcode={500046}, 
            state={Telangana},
            country={India}}
 \ead{roshanm@tifrh.res.in}
%% Abstract
\begin{abstract}
We provide a two dimensional deformation model to describe how soft squishy circular particles respond to external forces and collisions. This model involves formulating mathematical equations and algorithms for the shape of a deformed particle based on the geometrical as well as physical properties of the particles, such as their size and elasticity.
The model also accounts for the conservation of energy associated with the particle shape during interaction with other particles as well as with rigid walls. To verify the accuracy of the shape of the deformed particles, we start with a single deformable particle in an overcompressed box which we compare with the numerically simulated deformable polygon (DP) model. Then we extend our method to a jammed packing of deformable particles which we also verify through the DP model. We also provide a protocol to calculate the total force acting on such deformable particles solely based on their deformation and to obtain both athermal jammed packings as well as packings in thermal equilibrium. 
\end{abstract}

%\includegraphics[scale=0.25]{zoomed_DP_plus_thoery.png}


%% Keywords

\end{frontmatter}



\section{Introduction}
Deformable particles refer to macroscopic elastic particles that can undergo deformations such as stretching, bending, and twisting in response to external forces or interactions with other particles. Examples of deformable particles include foams~\cite{bolton1991effects}, bubbles, emulsions~\cite{brujic2003measuring}, and biological cells and tissues~\cite{angelini2011glass,sadati2013collective,herremans2015automatic}. Efficient simulation of deformable particles is crucial in various fields, including computer graphics, engineering, biological systems, and materials science.
Unlike systems made up of hard, rigid particles where the arrangement of particles determines the mechanical properties, in soft granular systems, the deformability of the particles significantly influences the mechanical properties.
Simulating deformable particles efficiently is complex and computationally intensive, especially with large numbers of particles. Many numerical studies on athermal solids simplify deformable particles as spheres interacting via purely repulsive pairwise potentials, such as Hertzian, Harmonic, or the repulsive part of the Lennard-Jones model~\cite{durian1995foam,o2002random,o2003jamming,maharana2022athermal,maharana2022first,roshan_universal_2024,Stress_correlations_crystal}. These models often neglect the change in particle shape during interaction, considering the overlap ($\delta$) between particles as effective deformation. This simplification limits the applicability of soft potential models to real-life situations.
Soft granular systems are often modeled by Hertzian potential, where the normal contact force between two elastic spherical beads is given by Hertz contact force, $F \propto \delta^{3/2}$ where $\delta$ is the overlap~\cite{landau1987theoretical}. However, this approximation is only valid for small deformations within the elastic limit. Experiments on elastic homogeneous soft cylinders have shown different non-linear behavior at higher values of deformation~\cite{dagro2019nonlinear}. 

There have also been several studies that try to incorporate the shape of the deformed particle into the model. For instance, studies on foams consider the shape of the foam during deformation~\cite{bolton1991effects,bolton1992effects,winkelmann20172d}, where the bubbles are generated using PLAT software~\cite{bolton1996software}. However, the bubbles generated by PLAT show different structural properties compared to well-studied soft particle models. For example, in bubbles, the variation of the excess coordination number ($\delta z$) with excess packing fraction ($\delta \phi$) above jamming show linear behavior whereas soft particles models show a square-root variation~\cite{winkelmann20172d}. There are also confluent models, such as the Vertex or Voronoi models, extensively used in biological systems such as epithelial tissues and the eye facets of Drosophila~\cite{barton2017active,huang2023bridging,kim2016hexagonal}. The non-confluent vertex model is also used to study the dynamics of embryonic tissues~\cite{kim2021embryonic}.

Only a few numerical approaches for modeling deformable particles have been developed over the past two decades, with most of them being based on either the discrete element method (DEM) or the finite element method (FEM) ~\cite{nezamabadi2017modeling,mollon2018mixtures,cantor2020compaction,mollon2022soft}.
One such DEM model is the deformable polygon (DP) model, which is effective in both lower packing fractions as well as confluent systems~\cite{boromand2018jamming}. The DP model accounts for particle shape during deformation and successfully reproduces the structural properties observed near the jamming transition~\cite{boromand2019role}. It is widely used in the study of biological systems and soft granular packings~\cite{treado2022localized,cheng2022hopper}. Even though it is an elegant model, it is computationally expensive as each deformable particle is defined by several degrees of freedom. The aim of this article is to reduce the computational cost of creating deformable particles with properties similar to those of the DP model using a geometric approach. Our model significantly reduces the number of degrees of freedom, making it more accessible for larger system sizes and potentially faster for both static packings through energy minimization and for studying dynamical properties at finite temperatures.

In Section~\ref{sec_model}, we present a brief overview of the DP model. In Section~\ref{sec_geometry_deformable}, we introduce a geometric method for determining the shape of a deformable particle by following certain assumptions on the deformed particle shape. In Section~\ref{sec_onep_deformation}, we verify the accuracy of the analytical shape of a single deformable particle compressed by polygonal confinements by comparing it with an equivalent deformable polygon. In Section~\ref{sec_multip_deformation}, we extend our methods to jammed packing multiple deformable particles in a rectangular confinement. We provide a comparison between the numerically generated packing of deformable polygons and the corresponding packing generated from the center of mass and radius of the particles using the geometric shape functions defined in this study. In Section~\ref{sec_protocal_update}, we provide a possible protocol for obtaining the force on each particle and particle position update. Finally, in Section~\ref{sec_conclusion}, we conclude by suggesting possible limitations and future directions.
%%%%%%%%%%%%%%%%%%%%%%%%%%%%%%%%%%%%%%%%%%%%%%%%%%%%%%%%%

\section{Deformable polygon model}
\label{sec_model}
In this study, we derive a theoretical form for the shape function of a defomable particle and use the deformable polygon model as a template to verify our theoretical description.
In the deformable polygon model~\cite{boromand2018jamming,boromand2019role}, the deformable particle is represented by a  circular ring of soft beads connected by springs where any deviation from the circular shape has the following energetic cost:
\begin{figure*}[t!]
\centering
\includegraphics[scale=0.59]{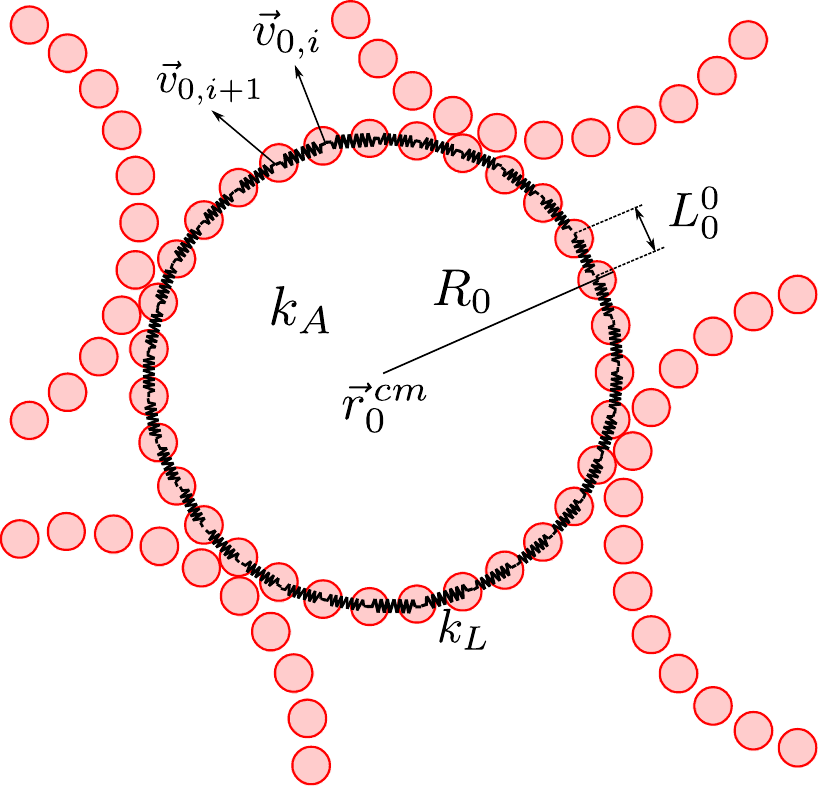}
\caption{Schematic of a deformable polygon with $N_v=31$.}
\label{fig_deformable_polygon}
\end{figure*}
\begin{equation} 
\begin{aligned}
        &U=\frac{k_L}{2}
        \sum_{m=1}^{N}\sum_{i}^{N_v}(L_{m,i}-L_m^0)^2+\frac{k_A}{2}\sum_{m}^{N}(A_{m}-A_m^0)^2+U_{int},\\
        &U_{int}=\frac{k_r}{2}\sum_{m=1}^{N}\sum_{n>m}^{N}\sum_{i,j=1}^{N_v}\left(2d-|\vec{v}_{m,i}-\Vec{v}_{n,j}|\right)^2 \times\Theta\left(2d-|\vec{v}_{m,i}-\Vec{v}_{n,j}|\right),
        \end{aligned}
        \label{eq_full_int}
\end{equation}
where $k_L$ is the spring constant between two consecutive beads in the ring and $L^0_m$ is the rest length of the springs in the $m$-th ring whereas the $L_{m,i}$ is the $i$-th spring length in the $m$-th ring after the deformation. Similarly, $k_A$ is the spring constant associated with the area fluctuation with respect to the undeformed ring. $N_v$ is the number of beads (or vertices) on each ring and $N$ is the number of DP particles in the system. Since an undeformed ring is a regular polygon, we can obtain the rest length between two vertices and the area of the DP particle as $L_m^0=2R_0\sin{(\pi/N_v)}$ and $A_m^0=N_vR_0^2\sin{(\pi/N_v)}$ respectively. In a deformed ring, we can obtain $L_{m,i}$ by measuring the distance between the vertex $\vec{v}_{m,i}$ and $\vec{v}_{m,i+1}$ and $A_m$ using the Surveyor’s Area Formula (Shoelace Formula)~\cite{braden1986surveyor}. Here $U_{int}$ is the interaction energy between two neighboring DP particles which is calculated by measuring the overlap between beads from the neighboring particles. Here each bead situated on the vertices of the ring is considered as a harmonically repulsive circular particle of radius $d$.

\section{Geometry of a deformed particle}
\label{sec_geometry_deformable}
Let us consider a two dimensional system of circular deformable particles of radius $R_m$ where the center of mass of the undeformed particle can be represented by $\vec{r}_m^{\,cm}$. The locus of the outer perimeter of the deformable particle with respect to this center can be parameterised as
\begin{equation}
    \begin{aligned}       x_m(\theta)=&x_m^{\,cm}+g_m(\theta)\cos{\theta},\\       y_m(\theta)=&y_m^{\,cm}+g_m(\theta)\sin{\theta},
    \end{aligned}
\end{equation}
where $x_m^{\,cm},y_m^{\,cm}$ correspond to the Cartesian coordinates of the particle center in the undeformed state. The aim of this article is to derive the shape function $g(\theta)$ given the particle positions and their radii. When the particle is undeformed, it assumes a circular shape to minimize its area, resulting in $g_m(\theta) = R_m$. To determine the shape function  in the deformed state, we make the following assumptions:

\begin{itemize}
    \item The contact region between two neighboring particles is a straight line segment ($l$) whose length depends on the distance between the neighbors as well as the magnitudes of the parameters $k_L$ (perimeter stiffness) and $k_A$ (area stiffness).
    \item  In the area between two contact regions, the particle forms a circular arc that is tangent to both contact line segments.
   
    \item We consider fully deformable particles with no overlap between neighboring particles. Consequently, the total potential energy of the system is determined solely by changes in the particle shapes, excluding any interaction potential $U_{int}$.
    
    \item Instead of a polygon, we treat each particle as a continuous deformable ring. Thus, the total potential energy can be expressed as a sum of an area contribution and a perimeter contribution:
    \begin{equation}
    \begin{aligned}
        &U= \sum_{m=1}^N (U^m_P +U^m_A),\\
        &U^m_P=\frac{k_P}{2}(P_{m}-P^0_m)^2,\text{ and, }U^m_A=\frac{k_A}{2}(A_{m}-A^0_m)^2,\\
        \end{aligned}
        \label{eq_area_energy}
    \end{equation}
    where $A^0_m=\pi R_m^2$ and $P_m^0=2\pi R_m$ are the initial area and perimeter of the undeformed particle, respectively, and $k_P=k_L/N_v$. The perimeter and area of the deformed particle can be expressed as:
    \begin{equation}
    \begin{aligned}
         P_m=&\frac{1}{2}\int_0^{2\pi}\sqrt{g_m^2+\left(\frac{\partial g_m}{\partial\theta}\right)^2}\mathrm{d}\theta,\\
         A_m=&\frac{1}{2}\int_0^{2\pi}g_m^2\text{d}\theta.
    \end{aligned}   
    \end{equation} 
    
    \item We assume that the deformable particles do not slip on top of each other to mimic the behaviour of the frictional deformable polygons.
\end{itemize}

\begin{figure*}[t!]
\centering
\includegraphics[scale=0.19]{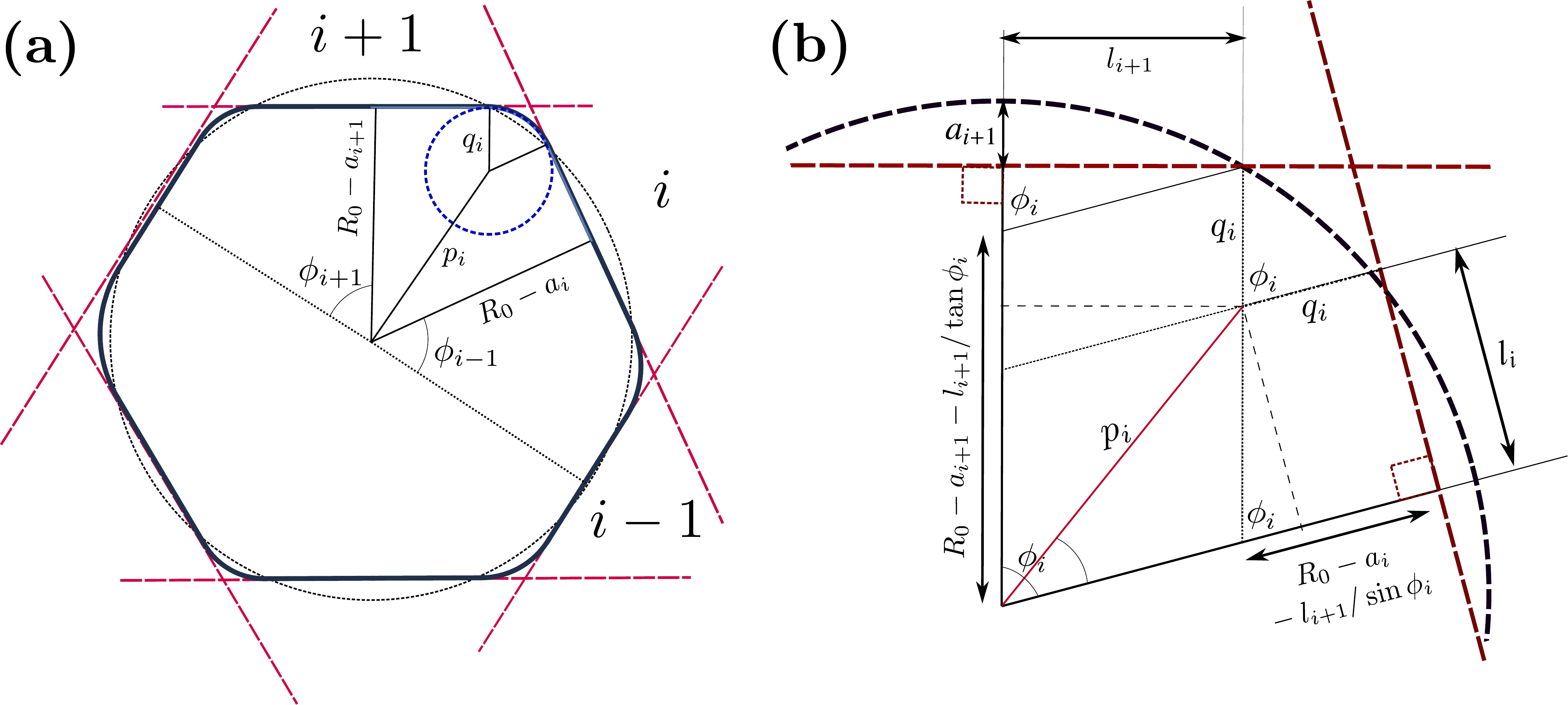}
\caption{\textbf{(a)} Deformation of a single DP particle of initial radius $R_0$ by linear surfaces represented by the dashed red line. Here the particle shape is obtained through FIRE energy minimization of the shape-energy function given in Eq.~\eqref{eq_full_int}. The angle between the contact $i$ and $i+1$ is represented by $\phi_i=\theta_{i+1}-\theta_i$, the distance from the center of the particle to the center of the circular arc between contacts $i$ and $i+1$ is represented by $p_i$ and $q_i$ is the radius of the circular arc. \textbf{(b)} The schematic of the particle and boundary between contacts $i$ and $i+1$. Here $l_i$ and $l_{i+1}$ correspond to the lengths of the contact regions between the particle and its neighbors $i$ and $i+1$ respectively. }
\label{fig_single_deformation_schematic}
\end{figure*}

Under these constraints, we consider a single deformable particle of undeformed radius $R_0$ ($m = 0$) situated at $\vec{r}_0^{\,cm}$ which is surrounded by $z$ neighbors with radii $R_i$ and positioned at $\vec{r}_i^{\,cm}$ where $i$ ranges from $1$ to $z$. The condition for a particle $i$ to be a neighbor can be represented as: $\left|\vec{r}_i^{\,cm}-\vec{r}_0^{\,cm}\right| \leq (R_i+R_0)$. The distance between the particle centre $\vec{r}_0^{\,cm}$ and the contact surface with the $i$-th neighbor can be written as $R_0-a_i$ where,

\begin{equation}
a_i=\frac{R_i^2-\left(\left|\vec{r}_i^{\,cm}-\vec{r}_0^{\,cm}\right| -R_0\right)^2}{2\left|\vec{r}_i^{\,cm}-\vec{r}_0^{\,cm}\right|}.
\label{eq_ai}
\end{equation}

The angle between two consecutive contacts ($i$ and $i+1$) (see Fig.~\ref{fig_single_deformation_schematic}) can be written as,
\begin{equation}
    \phi_i= \underbrace{\tan^{-1}\left(\frac{y^{\,cm}_{i+1}-y^{\,cm}_0}{x^{\,cm}_{i+1}-x^{\,cm}_0}\right)}_{\theta_{i+1}}-\underbrace{\tan^{-1}\left(\frac{y^{\,cm}_{i}-y^{\,cm}_0}{x^{\,cm}_{i}-x^{\,cm}_0}\right)}_{\theta_{i}}.
    \label{eq_phii}
\end{equation}

Let $l_i$ be the part of the perimeter of the particle that is in contact with the $i$-th neighbor. Then the region between two such consecutive contacts can be represented by two line segments of length $l_i$, $l_{i+1}$ and a circular arc with a radius $q_i$ and an arclength of $\phi_i q_i$. To derive a relation between these quantities we compare the energy associated with area fluctuation to the energy associated with the perimeter fluctuation between two such contact regions. For a fixed value of $k_P$ and $k_A$, we can write the energy change due to perimeter fluctuation and area fluctuation between two contact regions as
\begin{equation}
    \begin{aligned}
        (U_P^0)^{i,i+1}=&\frac{k_P}{2}(l_i+l_{i+1}+q_i\phi_i-R_0\phi_i)^2,\\
        (U_A^0)^{i,i+1}=&\frac{k_A}{2}\left[(R_0+q_i-a_i)l_i+(R_0+q_i-a_{i+1})l_{i+1}+\phi_iq_i^2-\phi_iR_0^2\right]^2.
    \end{aligned}
\end{equation}

In our study, we consider $k_P \gg k_A$, which implies a very high energy cost for any change in the particle perimeter compared to energy cost for change in the particle area. Thus, the major contribution to deformation comes from the area fluctuation. Considering the non-slip contact constraint, we can compare the change in the energy due to area ($U_P^{i,i+1}$) and perimeter ($U_A^{i,i+1}$) change between two contacts for $k_A/k_P \to 0$, which gives
\begin{equation} 
l_i+l_{i+1}+q_i\phi_i=R_0\phi_i.
\label{eq_l12_q}
\end{equation}

This relation indicates that the perimeter of the deformable particle between two contacts remains constant before and after the deformation.
We can also obtain an additional relationship between $l_i,l_{i+1}$ and $q_i$ that is independent of Eq.~\eqref{eq_l12_q}, by using the geometric constraint that the circular arc is tangent to both the contact line segments (see Fig.~\ref{fig_single_deformation_schematic}), which can be represented as:
\begin{equation}
\begin{aligned}
    &q_i=R_0-\frac{a_i+a_{i+1}}{2}-\frac{l_i+l_{i+1}}{2}\cot{(\phi_i/2)}, \\
    &l_{i+1}=l_i+\frac{a_{i+1}-a_i}{\tan{(\phi_i/2)}}.
\end{aligned}
\label{eq_q_i}
\end{equation}

A detailed derivation of the above geometric relations are given in the~\ref{app1}. Next, using the perimeter constraint in Eq.~\eqref{eq_l12_q} and the geometric constrain in Eq.~\eqref{eq_q_i}, we can obtain,
\begin{equation}
    \begin{aligned}
        l_{i}&=\frac{(a_{i}-a_{i+1})\sin{\phi_i}-(a_{i}\cos{\phi_i}-a_{i+1})\phi_i}{2-2\cos{\phi_i}-\phi_i\sin{\phi_i}},\\
        l_{i+1}&=\frac{(a_{i+1}-a_{i})\sin{\phi_i}-(a_{i+1}\cos{\phi_i}-a_{i})\phi_i}{2-2\cos{\phi_i}-\phi_i\sin{\phi_i}},\\
        q_i&=R_0-(a_i+a_{i+1})\frac{1-\cos{\phi_i}}{2-2\cos{\phi_i}-\phi_i\sin{\phi_i}}\text{,    }\forall \phi_i \in \{0,2\pi\}.
    \end{aligned}
    \label{eq_l1_l2_qi}
\end{equation}

The position of the center of the circular arc with respect to the particle center $\vec{r}_{0}^{\,cm}$ can be represented as 
\vspace{0.1cm}
\begin{equation}
    \begin{aligned}       &x^q_i=x_0^{\,cm}+p_i\cos{\theta_i^p} \text{, and, }y^q_i=y_0^{\,cm}+p_i\sin{\theta_i^p},\\
    &\text{where, }p_i=\sqrt{l_i^2+(R_0-a_i-q_i)^2},\\
    &\hspace{1.1cm} \theta_i^p=\theta_i+\tan^{-1}\left(\frac{l_i}{R_0-a_i-q_i}\right).
    \end{aligned}
    \label{eq_pi}
\end{equation}

Next using the formulae given in Eqs.~\eqref{eq_ai},~\eqref{eq_phii}, ~\eqref{eq_l1_l2_qi} and ~\eqref{eq_pi}, we can obtain the functional form of the contour of the deformed particle situated at $\vec{r}_0^{\,cm}$ in polar coordinate system and the angular variation of  
 the $g_j (\theta)$ can be written as:

\begin{equation} 
\begin{aligned}      
g_0(\theta)=\sum_{i=1}^{z}&\left[\frac{(R_0-a_i)}{\cos{(\theta-\theta_{i})}}H\left((\theta-\theta_{i})(\theta^s_{i}-\theta)\right)+\frac{(R_0-a_{i+1})}{\cos{(\theta_{i+1}-\theta)}}H\left((\theta-\theta_{i}^e)(\theta_{i+1}-\theta)\right)\right.\\
&\left.+\left(p_i\cos{\left(\theta^p_i-\theta\right)}+\sqrt{q_i^2-(p_i\sin{\left(\theta^p_i-\theta\right)})^2}\right)H\left((\theta-\theta^s_{i})(\theta^e_{i}-\theta)\right)\right],
        \end{aligned}
        \label{eq_def_particle_shape}
\end{equation}
where $H$ corresponds to the Heaviside theta function.  $\theta_i^s$ and $\theta_i^e$ correspond to the start and end polar angle of the circular arc region with respect to the particle center $\vec{r}_0^{\,cm}$ and can be represented as
\begin{equation}
    \begin{aligned}
\theta^s_i=&\theta_{i}+\tan^{-1}\left(\frac{l_i}{R-a_{i}}\right),\\
      \theta_i^e=&\theta_{i+1}-\tan^{-1}\left(\frac{l_{i+1}}{R-a_{i+1}}\right).
    \end{aligned}
\end{equation}

In an undeformed particle, $a_i=l_i=p_i=0$, which makes $g_0(\theta)=q_i=R_0$. Here, the particle shape is determined using the constraint that the perimeter of the particle is always constant at any level of deformation. However, different types of constraints can also be used, such as fixed area deformable particles or mixed situations where both the perimeter and the area are allowed to fluctuate. For instance, considering deformable particles where $K_A \gg K_P$, we can obtain a different constraint on the particle shape i.e. 
\begin{equation}
    (R_0+q_i-a_i)l_i+(R_0+q_i-a_{i+1})l_{i+1}+\phi_iq_i^2=\phi_iR_0^2.
    \label{eq_constraint_area}
\end{equation}

The above constraint corresponds to a situation where the particle can change its shape to fit within the confines of the surrounding geometry while maintaing a constant area during deformation.
Therefore, by using the constraint for constant area of the particle as given in Eq.~\eqref{eq_constraint_area} and the geometrical constraints for the contact region and the circular arc described in Eq.~\eqref{eq_q_i}, we can obtain the parameters associated with the shape of the deformed particle. For example, we can obtain the lengths of the contact regions within an edge between two consecutive contacts $i$ and $i+1$ as,

\begin{equation}
\begin{aligned}
    l_i&=-(R_0-a_i) \cot \phi_i -(R_0-a_{i+1}) \csc \phi_i -\frac{\phi_i ^2}{4}+\\
    &\frac{ \sqrt{\phi_i ^4\cos^2 (\phi_i/2)  -4 R_0 \phi_i ^2 \sin \phi_i-4 \cos \phi_i\left((R_0-a_i)^2+(R_0-a_{i+1})^2\right)+8 (R_0-a_{i}) (R_0-a_{i+1})
   }}{4 \cos \left(\phi_i /2\right)},
\end{aligned} 
\end{equation}
where $a_i,\phi_i$ can be obtained using Eq.~\eqref{eq_ai} and Eq.~\eqref{eq_phii}. For any values of $k_P$ and $k_A$, the given functional forms of $a_i,\phi_i,p_i,\theta_i^p$ are all same. The only differece lies in the contact length and the radius of the circular arc which varies with the different stiffness constants.

\section{Single deformable particle in an overcompressed polygonal box}\label{sec_onep_deformation}

\begin{figure}[t!]
\centering
\includegraphics[scale=0.12]{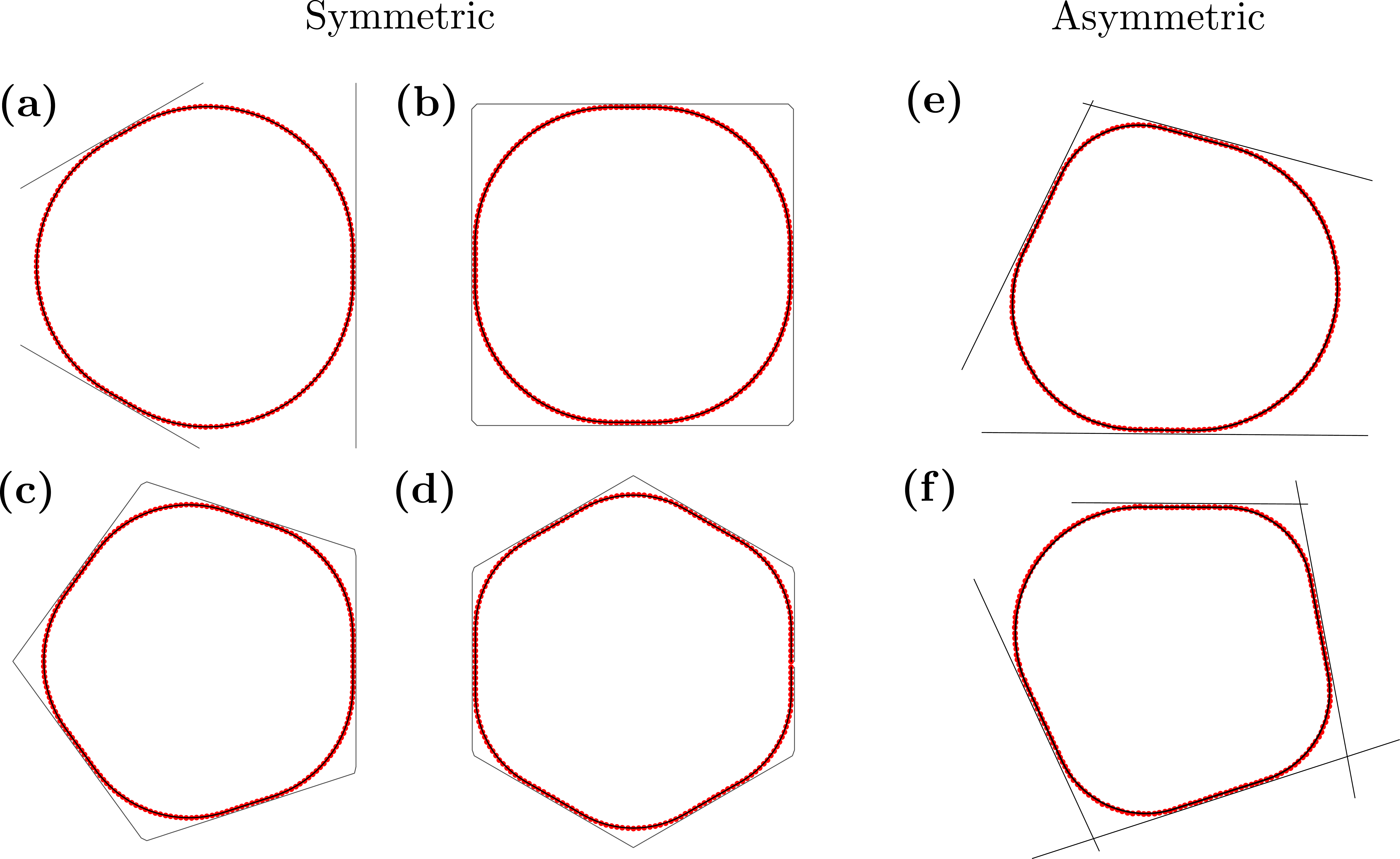}
\caption{Comparison of shape of a single deformable particle in boxes of both symmetrical and asymmetric polygon geometric shapes obtained from numerically simulated DP particle with $188$ beads and the constrained geometry shape $g_0(\theta)$. The red points represent the beads of the deformed DP particle, while the solid black line represents the constrained geometric shape of the deformed particle given by Eq.~\eqref{eq_def_particle_shape}. Figures \textbf{(a)}, \textbf{(b)}, \textbf{(c)}, and \textbf{(d)} correspond to regular polygonal confinements with $n=3, 4, 5,$ and $6$ sides, respectively. Figures \textbf{(e)} and \textbf{(f)} correspond to asymmetric polygonal confinements with $n=3$ and $4$ sides, respectively.}
\label{fig_one_particle_def_compare}
\end{figure}

To verify the validity of the shape function defined in Eq.~\eqref{eq_def_particle_shape}, we perform simulations with a single deformable particle modeled using a spring-mass ring system, as detailed in Section \ref{sec_model}. We place this particle within a closed confinement shaped as a $n$-sided polygon (with $n$ chosen to be 3, 4, 5, or 6). Initially, the particle is positioned at the center of the confinement, with the distance from the center to each side of the polygon greater than the radius of the particle. We then isotropically compress the polygonal box. By the end of the compression, the distance from the box sides to the center is less than the radius of the particle, causing the its shape to deform gradually. Here we assume non-slip boundary to mimic interaction between two deformable polygons i.e. the DP is not allowed to slip on the surface of the walls. The final shape of the deformable particle is achieved by minimizing the total shape energy using the FIRE  algorithm~\cite{bitzek2006structural}. Throughout the simulation, we maintain the spring constant $k_A$ at $1$. We vary the spring constant $k_L$ and the degree of overcompression, denoted by $a$. For each value of $k_L$, we incrementally increase the compression by steps of $0.005$ (i.e., $a/R_0=0.005, 0.01, ..., 0.05$), and minimize the energy at each step to obtain the final shape of the particle. The energy-minimized shape in the deformable polygon (DP) model is represented by the bead positions $\vec{v}_{0,i}$. In Fig.~\ref{fig_one_particle_def_compare}, we have presented the shape of a DP particle compressed by polygonal boxes with different numbers of sides. We have also plotted the theoretical shape of these particles defined in the following paragraph on top of the DP particles (denoted by black lines in Fig.~\ref{fig_one_particle_def_compare}).

\begin{figure*}[t!]
\centering
\includegraphics[scale=0.25]{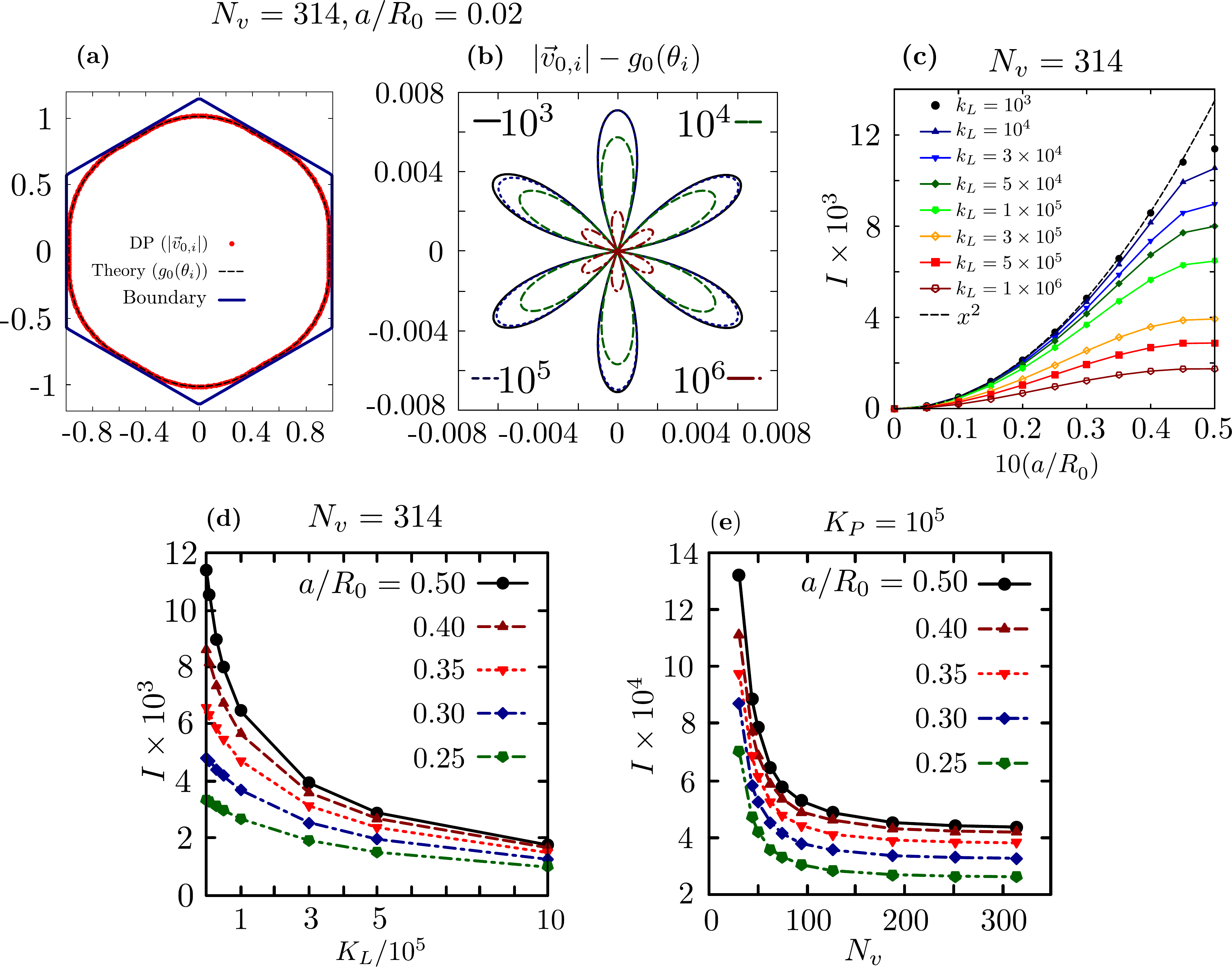}
\caption{\textbf{(a)} A deformable polygon (DP) particle with an initial radius of $R_0 = 1$ is placed inside a hexagonal box. The perpendicular distance from the origin to any side of the hexagon is $R_0 - a$, where $a/R_0 = 0.02$. The red points represent the beads on the DP ring, and the dashed black line depicts the theoretical shape of the deformed particle.
\textbf{(b)} This plot shows the difference between the theoretical shape of the deformed particle and the shape obtained numerically using the DP model. Different lines correspond to various values of $k_L$. The difference ($\left|\vec{v}_{0,i}\right| - g_0(\theta_i)$) decreases gradually as $k_L$ increases in the DP model.
\textbf{(c)} For each value of compression ($a$) by the boundary walls, the net difference between the theoretical shape and the DP model is measured by calculating the summation $I = N_v^{-1}\sum_{i=1}^{N_v} \left|\left|\vec{v}_{0,i}\right| - g_0(\theta_i)\right|$. This calculation is repeated in the DP model for different values of $k_L$, with $k_A = 1$ and $N_v = 314$ kept constant across all simulations.
\textbf{(d)} The net difference $I$ is measured as a function of spring constant $k_L$ for different magnitudes of overcompression. \textbf{(e)}  $I$ measured as a function of the number of beads on the DP for different magnitudes of overcompression. Here spring constant associated with the perimeter is kept constant, resulting in  $K_L=N_vK_P$, for different $N_v$.}
\label{fig_onep_full}
\end{figure*}
Below we give the expression for the shape of the deformed particle in a regular polygonal confinement. Given the center of mass of the particle, we express the shape function as outlined in Eq.~\eqref{eq_def_particle_shape}. This shape function is designed to describe the deformed geometry of the particle as it adapts to the constraints of the surrounding confinement.
Let $R'$ denote the perpendicular distance from any side of the regular polygonal confinement to the center of the polygon, and $R_0$ denote the radius of the particle when it is in an undeformed, relaxed state. As we compress the confinement, the distance $R'$ becomes smaller than $R_0$, forcing the particle to deform to fit within the new boundaries. To understand how the particle shape changes in response to this compression, we leverage the discrete rotational symmetry of the confinement. Regular polygons, by definition, have sides and angles that are uniformly distributed. For instance, in a regular $n$-sided polygon, the angle between any two adjacent sides is $2\pi/n$. This symmetry allows us to focus on a single section of the particle, knowing that the behavior in this section will be replicated around the entire perimeter. Using this symmetry we define the contact regions between the particle and the confinement as:
\begin{equation}
    \begin{aligned}
        a_i&=R_0-R', \hspace{0.5cm} \phi_i=2\pi/n,\hspace{0.5cm} \theta_i=2(i-1)\pi/n,\\
        l_{i}&=\frac{(R_0-R')(1-\cos{(2\pi/n)})(2\pi/n)}{2-2\cos{(2\pi/n)}-(2\pi/n)\sin{(2\pi/n)}}.
    \end{aligned}
    \label{eq_onep_para_1}
\end{equation}

Similarly, other parameters such as the position and radius of the deformed circular arc can also be obtained as
\begin{equation}
    \begin{aligned}
        q_i&=R_0-\frac{2(R_0-R')(1-\cos{(2\pi/n)})}{2-2\cos{(2\pi/n)}-(2\pi/n)\sin{(2\pi/n)}},\\
        p_i&=\frac{(R_0-R')}{\cos{(\pi/n)}-(n/\pi)\sin{(\pi/n)}},\text{ and, }
        \theta_i^p=\frac{(2i-1)\pi }{n}.       
    \end{aligned}
     \label{eq_onep_para_2}
\end{equation}

Now, using the parameters defined in Eq.~\eqref{eq_onep_para_1} and Eq.~\eqref{eq_onep_para_2}, we can define the shape of the deformed particle using Eq.~\eqref{eq_def_particle_shape}. Similarly, the expression for the shape of a deformed particle in an irregular confinement can also be obtained. In Fig.~\ref{fig_onep_full} \textbf{(a)}, we have shown the deformed particle obtained from simulation (represented by red points) and compared it with the perimeter-constrained geometric shape (represented by the black dashed line) for $k_L=10^{6}$, $N_v=314$ and overcompression, $a/R_0=0.02$. For a given bead position $\vec{v}_{0,i}$, we calculate the corresponding angle $\theta_i$ using the equation:
\begin{equation}
    \theta_i=\tan^{-1}\left(v_{0,i}^y/v_{0,i}^x\right).
\end{equation}

 This angle $\theta_i$ allows us to compare the simulated shape with the theoretical shape. Next, we measure the difference between the deformable polygon (DP) model and the theoretical shape as $\left|\vec{v}_{0,i}\right|-g_0(\theta_i)$. In Fig.~\ref{fig_onep_full} \textbf{(b)}, we have shown the difference between two models with increasing magnitude of $k_L$. We observe that as $k_L$  increases, the difference between the two models decreases, indicating that the theoretical model becomes a better approximation of the actual deformed shape. To numerically quantify this, we measure the net difference between the two shapes using the summation:
\begin{equation}
    I=\frac{1}{R_0N_v}\sum_{i=1}^{N_v} \left|\left|\vec{v}_{0,i}\right| - g_0(\theta_i)\right|.
\end{equation}

We measure $I$ for different values $k_{L}$ and observe that at higher values of compression, the analytical shape deviates from the DP shape more. However, $I$ decreases monotonically with $k_L$ for all the values of overcompression $a$. This suggest that for $k_A/k_L \to 0$, $I \to 0$ for all $a/R_0$.
This monotonic decrease in $I$ implies that the theoretical shape function $g_0(\theta)$, which is derived by considering the perimeter of the particle to be fixed, accurately predicts the shape of the deformable polygon when the stiffness ratio $k_A/k_L$ is very small. Therefore, for sufficiently large $k_L$, the analytical model $g_0(\theta)$ becomes an excellent approximation of the shape of a deformable particle confined in a polygonal box. We also measure the variation in $I$ with increasing number of beads on the DP for a fixed value of perimeter spring constant $k_P$ which is presented in Fig.~\ref{fig_one_particle_def_compare} (e). We observe that the $I$ decreases with increasing $N_v$ and eventually saturates to a fixed value for any levels of overcompression $a/R_0$. This saturation point is dependent on $k_P$, with higher values of $k_P$ leading to a lower saturation value of $I$. This outcome further validates the accuracy of the geometric method used to predict the shape of the DP, especially in the limit where $N_v \to \infty$ and $k_P \to \infty$ both approach infinity.

\section{Multiple deformable particle in an overcompressed rectangular box}\label{sec_multip_deformation}
To check the validity of this geometrical method for jammed packing of deformable polygons in a rectangular box, we consider a bidisperse system of soft particles where the ratio of the radius between the larger and smaller particles is $R^l/R^s=1.4$. We begin by distributing these soft particles randomly within a large square box. Subsequently, we compress the box isotropically while minimizing the total energy of the system using the FIRE algorithm. The compression is halted when the system becomes jammed, which is characterized by a finite positive stress. At this point, the particles are in close contact, and no further compression can occur without significantly increasing the energy of the system. Next, we replace each soft particle with a $N_v$-sided deformable polygon. The number of vertices for each particle is chosen based on its size i.e. the larger particle is assigned a greater number of vertices compared to the smaller particle, such that: 

\begin{equation}
    \frac{N^{l}_v}{N_v^{s}} \sim \frac{(R^l-p)}{(R^s-p)},
\end{equation}
where $N_v^l$ denotes the number of vertices on the larger particle, $N_v^s$ denotes the number of vertices on the smaller particle, $R^l$ is the radius of the larger particle, $R^s$ is the radius of the smaller particle, and $p$ is the radius of each bead on the vertices of the deformable polygons. We then minimize the energy of this packing where the energy is defined by Eq.~\eqref{eq_full_int}.

\begin{figure*}[t!]
\centering
\includegraphics[scale=0.25]{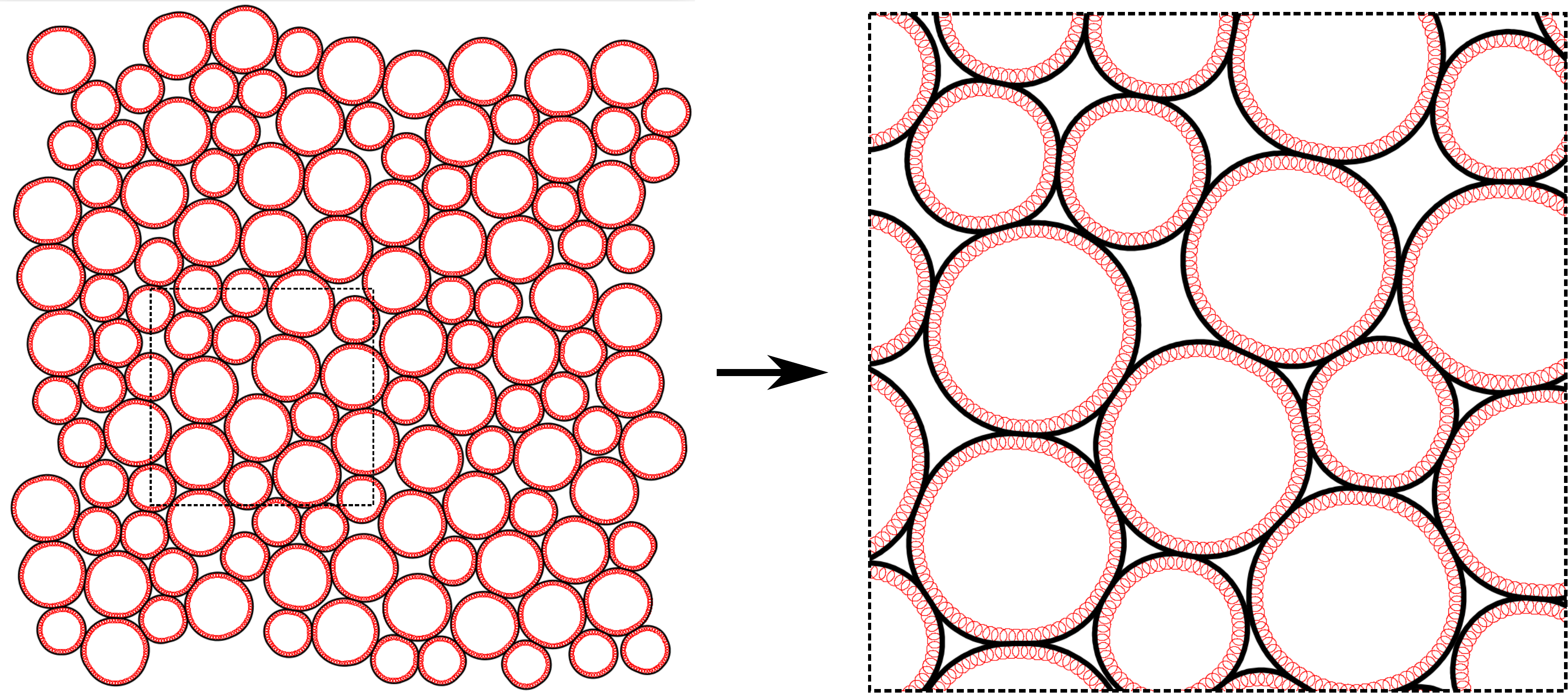}
\caption{Comparision of simulations of DP packing in an overcompressed box with the shape functions defined in Eq.~\eqref{eq_def_particle_shape}. The small red circles correspond to the beads of a numerically obtained deformable polygon whereas the solid black line corresponds to the theoretically obtained shape of the deformed particles given their center of mass and radius. A small region of the entire system has been zoomed in for better visibility of the DP beads.}
\label{fig_multip_def}
\end{figure*}
Next, we isotropically compress the system with an incremental compressive strain of magnitude $\delta \epsilon$. At each strain increment, the system is relaxed to its energy minimum, ensuring the protocol can be considered quasistatic.  At the jamming point, where the particles are in their undeformed states, each polygon can be approximated as a circular particle. However, as we further compress the system, the relaxation of the net energy occurs through two primary mechanisms: (1) particle rearrangements and (2) deformation of the particles. For this study, we have taken a system of $N=128$ deformable particles where the radii of smaller and larger particles are $R^l=0.7$ and $R^s=0.5$ respectively. We choose the radius of each bead on the vertex of the deformable polygons as $p=0.05$ resulting in a ratio of vertices $N_v^l/N_v^s=1.44$. Specifically, we select $N_v^l=72$ vertices for the larger particles and $N_v^s=50$ vertices for the smaller particles.

Next, we extend the geometrical method described in Section~\ref{sec_geometry_deformable} to multiparticle jammed packings within a rectangular box. Similar to a single particle in confinement, the contact region between two particles acts as a boundary. As we compress the box, the average distance between particles decreases, effectively bringing the boundary closer to the particle centers. We implement techniques detailed in Section~\ref{sec_geometry_deformable} to obtain the shape of the individual deformable particle in the system. However, the deformed particle shape can also overlap with other neighboring deformed particles which we refer as secondary overlap. These secondary overlaps are treated similarly to primary overlaps between two circles  with the radii of the circles being the radii of the circular arc. The same procedure is followed to resolve secondary deformations, and this process can be iterated for higher-order deformations. For most configurations the final deformed shape can be obtained just performing three iterations. In Fig.~\ref{fig_multip_def}, we have shown a jammed packing of deformable polygons (red circles) for an istropic strain above the jamming with an amplitude, $ \epsilon=0.01$. From this packing, we obtain the center of mass and radius of each particle. Using the extracted parameters and the constrained geometric shape method described by Eq.~\eqref{eq_def_particle_shape}, we generate the shapes of the particles in the packing. These shapes are represented by black solid lines in Fig.~\ref{fig_multip_def}.

\section{Protocol for particle position update}
\label{sec_protocal_update}

Using the functional form $g_0(\theta)$ in Eq.~\eqref{eq_def_particle_shape}, we can obtain the area of the deformed particle as:
\begin{equation}   \begin{aligned}A_{0}=\frac{1}{2}\sum_{i=1}^z&\left[\phi_i q_i^2+l_i(R_0-a_i+q_i)+l_{i+1}(R_0-a_{i+1}+q_i)\right].
    \end{aligned}
\end{equation}

Using the above formula for area we can determine the total energy ($U$) of the packing using Eq.~\eqref{eq_area_energy}. We can also determine the force on each deformable particle by taking the positional derivative ($-\nabla U$) of the total potential energy i.e.
\begin{equation}
\begin{aligned}
     f^x_0&=-K_A(A_0-A_0^0)\frac{\partial A_0}{\partial x_0^{\,cm}}-K_A\sum_{i=1}^z(A_i-A_i^0)\frac{\partial A_i}{\partial x_0^{\,cm}},\\
       f^y_0&=-K_A(A_0-A_0^0)\frac{\partial A_0}{\partial y_0^{\,cm}}-K_A\sum_{i=1}^z(A_i-A_i^0)\frac{\partial A_i}{\partial y_0^{\,cm}}.
\end{aligned}
\end{equation}

Since the potential energy only contains the area fluctuation term, the final positions of the particles in a jammed packing are highly dependent on their deformed area. However, other terms related to perimeter fluctuation and interaction between the neighbors can be included to force for higher values of $k_P$ and $k_r$.

The above formulation is invalid when the particle has only one neighbor as an edge can only be defined when coordination number $z \ge 2$. In such a scenario, we consider both the particles repel each other harmonically without deformation just like in soft particle systems via potential: $U_{ij}=\frac{k_A}{2}\left(1-\frac{r_{ij}^{\,cm}}{R_i+R_j}\right)^2$. 
We displace the particle in the direction normal to the contact in using the force $-\partial_{r_i}U_{ij}$ until it forms a new contact or the present contact breaks to form a rattler. Given the forces, we can update the particle positions at each time step using a verlet algorithm for dynamical systems and can achieve energy minimized packing through damped dynamics/ FIRE minimization.

\begin{figure}
    \centering
    \includegraphics[width=0.8\linewidth]{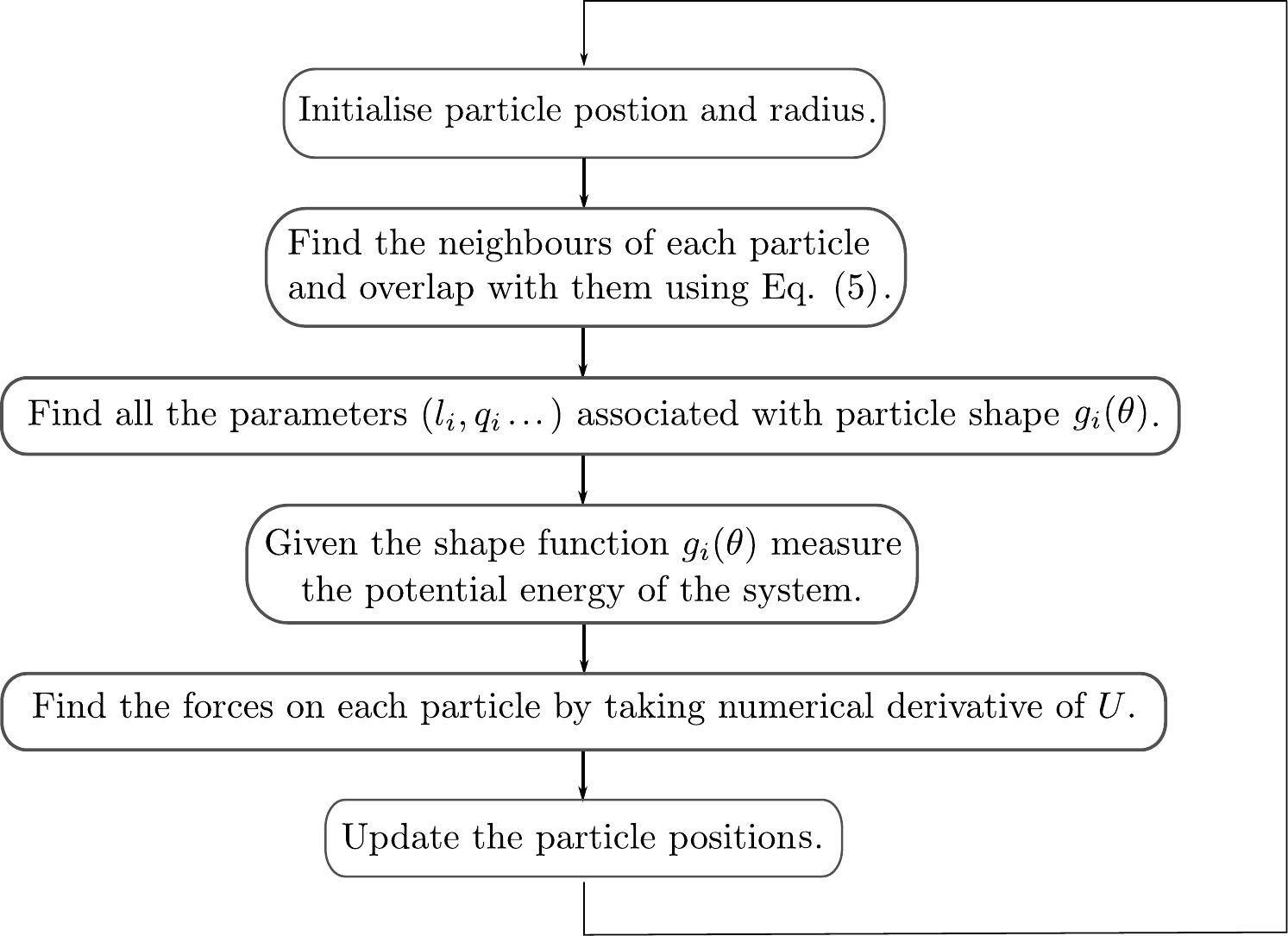}
    \caption{Protocol to simulate deformable particle packings.}
    \label{fig_flowchart}
\end{figure}

\section{Conclusion and Discussion}
\label{sec_conclusion}
In this study, we have proposed an efficient method to obtain jammed packings of deformable particles at higher packing fractions. Similar to soft particle models, our geometric formulation only requires the particle positions and their radii to determine the shape of individual particles after deformation, which in turn can be used to determine the total force acting on each particle. In the deformable polygon (DP) model, the total number of degrees of freedom in the system is $N_vN$, whereas our proposed model has only $N$ degrees of freedom for the same packing. This reduction in complexity makes our model more efficient and scalable, enabling us to create larger system packings more effectively.
By employing this geometric method to calculate the force between two deformable particles, we can develop a faster algorithm for generating static packings through energy minimization as well as for studying dynamical properties at finite temperatures.

One significant limitation of the proposed method arises when dealing with systems composed of multiple deformable particles with varying degrees of deformability. Specifically, if the particles in the system have different spring constants $k_P$ and $k_A$. The primary issue stems from the assumption that the contact regions between particles are perfectly flat. In reality, when deformable particles with different spring constants interact, the contact regions can exhibit complex shapes, including concave or convex curvatures, depending on the relative deformability of the particles. This geometric method, in its current form, does not account for these complex contact geometries, potentially leading to inaccuracies when predicting the behavior of systems with heterogeneous deformability. Despite this limitation, the proposed geometric approach can be extended to incorporate the shape of the contact regions, taking into account the specific deformabilities of the interacting particles which would enhance the accuracy of the method.

Using this geometric method, various structural and mechanical properties of the jammed packings can be studied and verified. For example, the average coordination number can be analyzed to understand the connectivity and stability of the packing. Additionally, the elastic moduli, which describe the stiffness of the material, can be measured above the jamming transition to explore how the mechanical response of the system changes with increased packing density. Our model also allows for direct studies on the effect of particle deformability on order-disorder transitions and amorphization transitions. Understanding these transitions is key to exploring how systems evolve from ordered to disordered states and how they become amorphous under varying conditions.

\section{Acknowledgements}
I would like to thank Kabir Ramola, Stephy Jose, Surajit Chakraborty, Rashmiranjan Bhutia, and Shilpa Prakash for useful discussions. This project was funded by intramural funds at TIFR Hyderabad from the Department of Atomic Energy (DAE). We acknowledge support of the Department of
Atomic Energy, Government of India, under
Project Identification No. RTI4007.
\appendix
\section{Derivation of geometric constraints}
\label{app1}

\textbf{(a) For $\phi_i<\pi$:}

In the right angle triangle \textbf{JIG}, we have $\mathbf{IG}=l_{i+1}$ and $\angle\mathbf{GJI}=\phi_i$. Using these, we can derive the following relationships:

\begin{equation}
    \mathbf{IJ}=l_{i+1}/\tan{\phi_i}, \text{ and, }\mathbf{JG}=\mathbf{OA}=l_{i+1}/\sin{\phi_i}.
\end{equation}

Next, we determine the length of $\mathbf{AG}$ as follows:
\begin{equation}
    \mathbf{AG}=\mathbf{OJ}=\mathbf{OH}-\mathbf{JI}-\mathbf{IH}=R_0-a_{i+1}-\frac{l_{i+1}}{\tan{\phi_i}}.
\end{equation}

We find the length of $\textbf{AF}$ by substracting $\mathbf{FG}$ from $\mathbf{AG}$:
\begin{equation}
    \mathbf{AF}=R_0-a_{i+1}-q_i-\frac{l_{i+1}}{\tan{\phi_i}}.
\end{equation}

In the right-angle triangle \textbf{ABF}, we can write $\mathbf{BF}=\mathbf{AF}\sin{\phi_i}$ and $\mathbf{AB}=\mathbf{AF}\cos{\phi_i}$. Given that, $q_i=\mathbf{FE}=\mathbf{BC}=\mathbf{OD}-\mathbf{OA}-\mathbf{AB}-\mathbf{CD}$, we can write: 
\begin{equation}
    \begin{aligned}
        q_i=&R_0-a_i-\frac{l_{i+1}}{\sin{\phi_i}}-\left(R_0-a_{i+1}-q_i-\frac{l_{i+1}}{\tan{\phi_i}}\right)\cos{\phi_i}\\
        q_i=&\frac{R_0-a_i-(R_0-a_{i+1})\cos{\phi_i}-l_{i+1}\sin{\phi_i}}{1-\cos{\phi_i}}.
    \end{aligned}
    \label{eq_qi_1}
\end{equation}

Using symmetry $q_i$ can also be represented as:
\begin{equation}
    \begin{aligned}
        q_i=&\frac{R_0-a_{i+1}-(R_0-a_{i})\cos{\phi_i}-l_{i}\sin{\phi_i}}{1-\cos{\phi_i}}.
    \end{aligned}
    \label{eq_qi_2}
\end{equation}

Thus $q_i$ can be represented as the average of Eq.~\eqref{eq_qi_1} and Eq.~\eqref{eq_qi_2},
\begin{equation}
    q_i=R_0-\frac{a_i+a_{i+1}}{2}-\frac{l_i+l_{i+1}}{2}\cot{\phi_i/2}.
\end{equation}

Similarly using $l_i=\mathbf{DE}=\mathbf{BF}$, we can write,
\begin{equation}
    \begin{aligned}
        l_i=&\left(R_0-a_{i+1}-q_i-\frac{l_{i+1}}{\tan{\phi_i}}\right)\sin{\phi_i},\\
        l_i=&l_{i+1}+(a_{i}-a_{i+1})\cot{\frac{\phi_i}{2}},
    \end{aligned}
\end{equation}
\begin{figure}
    \centering
    \includegraphics[width=0.95\linewidth]{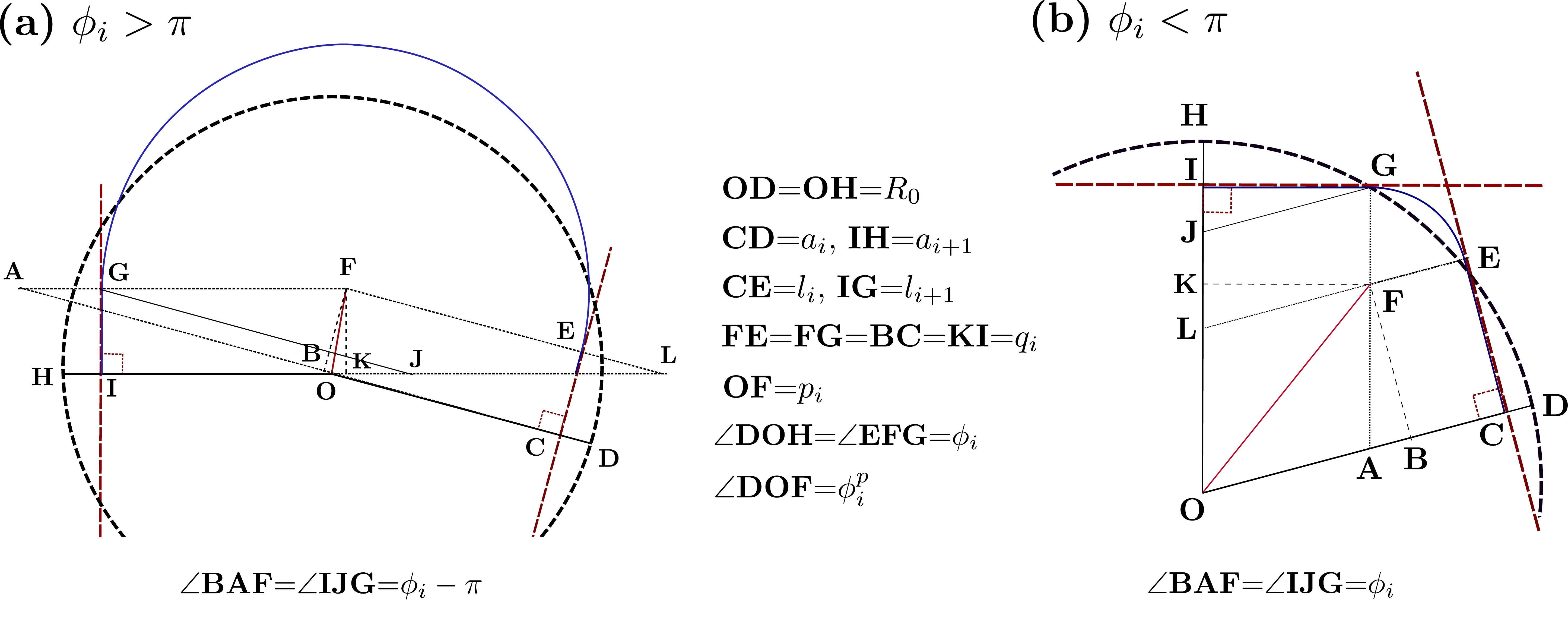}
    \caption{Geometry of deformation.  Here $\mathbf{OD}$ and $\mathbf{OH}$ are perpendiculars to the contact regions with the $i$-th and $i+1$-th neighbors respectively from the center of the circle. Here $\mathbf{CE}$ and $\mathbf{GI}$ are contact regions where the points $\mathbf{G}$ and $\mathbf{E}$ do not necessarily lie on the circle. $\mathbf{GJ}$ and $\mathbf{EL}$ are line segments drawn parallel to the $\mathbf{OD}$. Similarly, $\mathbf{GA}  \parallel  \mathbf{OH}$. }
    \label{fig_Geometry}
\end{figure}

\textbf{(b) For $\phi_i\ge \pi$:}

Using a similar approach, we obtain the following relationships:
\begin{equation}
    \begin{aligned}
        \mathbf{IJ}=&l_{i+1}/\tan{(\phi_i-\pi)}, \text{ and, }\mathbf{JG}=\mathbf{AO}=l_{i+1}/\sin{(\phi_i-\pi)},\\
        \mathbf{AG}=&\mathbf{OJ}=\mathbf{JI}-\mathbf{OH}+\mathbf{IH}=-R_0+a_{i+1}+\frac{l_{i+1}}{\tan{(\phi_i-\pi)}},\\
        \mathbf{AF}=&\mathbf{AG}+\mathbf{GF}=-R_0+a_{i+1}+\frac{l_{i+1}}{\tan{(\phi_i-\pi)}}+q_i,\\
        \mathbf{AB}=&\mathbf{AF}\cos{(\phi_i-\pi)}.
    \end{aligned}
\end{equation}

We can obtain the length of the radius of the circular arc as:
\begin{equation}
    \begin{aligned}
        q_i=&\mathbf{FE}=\mathbf{BC}=\mathbf{AO}-\mathbf{AB}+\mathbf{OD}-\mathbf{CD},\\
        q_i=&\frac{l_{i+1}}{\sin{(\phi_i-\pi)}}-\left(-R_0+a_{i+1}+\frac{l_{i+1}}{\tan{(\phi_i-\pi)}}+q_i\right)\cos{(\phi_i-\pi)}+R_0-a_i.
    \end{aligned}
\end{equation}

Simplifying the above equation and using symmetry we obtain,
\begin{equation}
    q_i=R_0-\frac{a_i+a_{i+1}}{2}-\frac{l_i+l_{i+1}}{2}\cot{\phi_i/2}.
\end{equation}

Similarly, the lengths of the contact regions are
\begin{equation}
    \begin{aligned}
        l_i=&\mathbf{CE}=\mathbf{BF}=\mathbf{AF}\sin{(\phi_i-\pi)},\\
        l_i=&l_{i+1}+(a_{i}-a_{i+1})\cot{\left(\frac{\phi_i}{2}\right)}.
    \end{aligned}
\end{equation}

  \bibliographystyle{elsarticle-num} 
  \bibliography{DP_Bib}

\end{document}